\documentclass[preprint, nofootinbib, aps, floatfix]{revtex4-1}

\usepackage[dvipsnames]{xcolor}
\usepackage{color}
\usepackage[latin1]{inputenc}
\usepackage{fontenc}
\usepackage{graphicx}
\usepackage{amsmath}
\usepackage{amssymb}
\usepackage{bbold}
\usepackage[margin=0.76in]{geometry}
\usepackage{xcolor}
\usepackage{hyperref}
\usepackage{cleveref}
\usepackage{soul}
\usepackage[caption=false]{subfig}

\newcommand{\p}{\partial}

\newcommand{\dg}{\dagger}

\renewcommand{\AA}{\mathcal{A}}
\newcommand{\DD}{\mathcal{D}}
\newcommand{\FF}{\mathcal{F}}
\newcommand{\JJ}{\mathcal{J}}
\newcommand{\OO}{\mathcal{O}}
\newcommand{\xperp}{\mathbf x_\perp}

\newcommand{\ev}[1]{\left< #1 \right>}
\newcommand{\tr}{\mathrm{Tr}}

\newcommand{\mbf}{\mathbf}
\newcommand{\GMV}{G}
\newcommand{\Gret}{G_\mathrm{ret}}

\usepackage{tcolorbox}

\begin{document}
	\count\footins = 1000 

	\title{Space-time structure of 3+1D color fields \\ in high energy nuclear collisions}
	\date{\today}
	\begin{abstract}
		We perform an analytic calculation of the color fields in heavy-ion collisions by considering the collision of longitudinally extended nuclei in the dilute limit of the Color Glass Condensate effective field theory of high-energy QCD. Based on general analytic expressions for the  color fields in the future light cone, we evaluate the rapidity profile of the transverse pressure within a simple specific model of the nuclear collision geometry and compare our results to 3+1D classical Yang-Mills simulations. 
	\end{abstract}
	
	\author{Andreas Ipp}
	\email{ipp@hep.itp.tuwien.ac.at}
	\affiliation{Institute for Theoretical Physics, TU Wien, A-1040 Vienna, Austria}
	
	\author{David I.~M\"uller}
	\email[Corresponding author: ]{dmueller@hep.itp.tuwien.ac.at}
	\affiliation{Institute for Theoretical Physics, TU Wien, A-1040 Vienna, Austria}
	
	\author{Soeren Schlichting}
	\email{sschlichting@physik.uni-bielefeld.de}
	\affiliation{Fakult\"at f\"ur Physik, Universit\"at Bielefeld, D-33615 Bielefeld, Germany}
	
	\author{Pragya Singh}
	\email{pragya@physik.uni-bielefeld.de}
	\affiliation{Fakult\"at f\"ur Physik, Universit\"at Bielefeld, D-33615 Bielefeld, Germany}
	
	\maketitle

	\section{Introduction}
	The space-time evolution of the Quark-Gluon-Plasma (QGP), produced in high-energy heavy-ion collisions (HIC) at the Large Hadron Collider (LHC) and Relativistic Heavy Ion Collider (RHIC), can be accurately described by relativistic viscous hydrodynamics \cite{Gale:2013da,Heinz:2013th}. However, the early stage of HICs, which provides the initial conditions for the subsequent hydrodynamic evolution, still requires a comprehensive understanding. Over the course of years, sophisticated pre-equilibrium models incorporating fluctuations at nucleonic and sub-nucleonic level \cite{Hirano:2005xf,Drescher:2007ax,Alver:2008aq,Schenke:2012hg,Mantysaari:2017cni} and frameworks including intermediate kinetic theory evolution \cite{Kurkela:2018wud,Kurkela:2018vqr} have been developed in order to obtain a complete theoretical description in quantitative agreement with the experimental observations. Even though these models have played a significant role to improve the understanding of the initial state and properly characterize the subsequent QGP dynamics~\cite{Bernhard:2015hxa,Bernhard:2016tnd,Nijs:2020roc}, the current state of the art modeling is often carried out at the level of an effectively 2+1D boost-invariant description and tends to ignore the longitudinal dynamics of heavy-ion collisions.
	
	The development of initial state models has benefited from first principles insights into the initial stage of heavy-ion collisions provided by the Color Glass Condensate (CGC) effective theory of high-energy QCD \cite{Gelis:2010nm,Gelis:2012ri}, where small $x$ partons in high energy nuclei are described by classical gluon fields, whose dynamics is governed by the classical Yang-Mills (CYM) equations. On the one hand, the CGC framework has lead to the development of various initial state models such as IP-Glasma \cite{Schenke:2012wb,Schenke:2012hg} and MC-KLN \cite{Drescher:2006ca,Drescher:2007ax}, which along with relativistic viscous hydrodynamics have been successful in describing azimuthal anisotropies  and charged hadron multiplicity \cite{Qiu:2011fi,Qiu:2011hf,Gale:2012rq,Schenke:2011bn,Schenke:2012hg}. On the other hand, a plethora of (semi-)analytic calculations have been carried out within the CGC framework using expansions in color source densities \cite{Kovner:1995ts,Kovchegov:1997ke,Dumitru:2001ux,McLerran:2016snu} or near-field expansions \cite{Fries:2005yc,Fukushima:2007yk,Fujii:2008km,Chen:2015wia,Carrington:2020ssh} in the boost-invariant limit, which have been further exploited to study the correlation function of the initial energy momentum tensor \cite{Dumitru:2001ux,Fujii:2008km,Lappi:2017skr,Albacete:2018bbv,Bhalerao:2019uzw,Guerrero-Rodriguez:2021ask} and jet momentum broadening in the early stages \cite{Ipp:2020mjc, Ipp:2020nfu}. Such numerical and (semi-)analytical results have also been important to guide the development of simple parametric initial state models such as IP-Jazma~\cite{Nagle:2018ybc} or TrENTo~\cite{Moreland:2014oya,Bernhard:2016tnd}, and to the development of a comprehensive understanding of the transverse dynamics of the fireball near mid-rapidity.

	Beyond the boost-invariant description of heavy-ion collisions, recent experimental studies, e.g.~of rapidity-dependent factorization breakdown \cite{PHOBOS:2006mfc,CMS:2015xmx,ATLAS:2017rij}, urge us to understand the dynamics of heavy-ion collisions beyond mid-rapidity, and have triggered an increased interest in the longitudinal structure of the initial state \cite{Ozonder:2013moa,Molnar:2014zha,Bozek:2015bna,Shen:2017bsr,Shen:2020jwv} and corrections to the eikonal limit \mbox{\cite{Altinoluk:2014oxa, Altinoluk:2015gia}}.
	While various implementations of 3+1D classical Yang-Mills equations have been developed either by varying the strengths of the classical sources \cite{Lappi:2004sf}, by generalizing the IP-Glasma model to 3+1D using JIMWLK rapidity evolution \cite{Schenke:2016ksl,McDonald:2018wql}, or by taking finite thickness of colliding nuclei into account \cite{Gelfand:2016yho,Ipp:2017lho, Ipp:2018hai,Ipp:2020igo,Schlichting:2020wrv}, we are not aware of any analytical calculation of the energy deposition in heavy-ion collision beyond the boost-invariant high-energy limit. 
	
	In this paper, we present the first analytical calculation of the initial energy deposition in heavy-ion collisions by solving the 3+1D classical Yang-Mills equations within the dilute limit of the Color Glass Condensate effective field theory of high-energy QCD. 
	Similar to the previous studies \cite{Gelfand:2016yho,Ipp:2017lho, Ipp:2018hai,Ipp:2020igo,Schlichting:2020wrv}, the longitudinal dynamics of the Glasma is analyzed by taking the finite extent of the colliding nuclei into account, while other corrections, such as the offset of the beam trajectory from the light cone, are neglected.
	We solve the linearized Yang-Mills equations and obtain analytic expressions for the perturbative gauge fields in the future light cone (analogous to the results of ~\cite{Kovner:1995ja,Dumitru:2001ux,McLerran:2016snu} for the boost-invariant case).
	Subsequently, we derive an analytic expression for the transverse pressure beyond the high-energy boost-invariant limit, for a specific realization of nuclear collision geometry. We establish the effectiveness of our perturbative calculation by comparing it to non-perturbative 3+1D classical Yang-Mills simulations \cite{Ipp:2018hai,Schlichting:2020wrv} for various thicknesses of the colliding nuclei. 
	
	This work is organized as follows: Starting in Section~\ref{Sec1}, we set up the formalism to study 3+1D collisions in the dilute limit and develop an auxiliary field approach to obtain the analytic expressions for the color fields produced in the future light cone. We then employ a simple model of nuclear collision geometry to derive analytic expressions for the transverse pressure in Section~\ref{AnalyticSect}
	and compare our (semi-)analytic results to 3+1D classical Yang-Mills simulation in Section~\ref{Sec3}. We conclude with Section~\ref{Sec4}.
	
	\section{General formalism}\label{Sec1}
	The Color Glass Condensate (CGC) framework provides an effective description of nucleus-nucleus collisions at high energies in terms of Yang-Mills theory. In the CGC formalism, hard partons of nuclei are modeled as color charges, which source soft partons in the form of classical color fields.
	Using light cone coordinates $x^\pm = (x^0 \pm x^3) / \sqrt{2}$, the color current of a nucleus moving along the negative $x^{3}$ direction (denoted as ``$A$'') is given by
	\begin{align}
		\JJ^\mu_A(x^+, \xperp) &= \delta^\mu_- \rho^a_A(x^+, \xperp) t^a, \label{eq:nuc_A_j}
	\end{align}
	where $\rho^a_A$ denotes the color charge density per unit transverse area, $\xperp = (x^1, x^2)$ denotes the transverse coordinates, $\delta^\mu_-$ is the Kronecker delta of the ``$-$'' light cone component and $t^a$ are the generators of the $SU(N_c)$ gauge group. The color current depends only on one of the two light cone coordinates (in this case $x^+$) and is assumed to be localized around $x^+ = 0$.
	The color field $\AA^\mu$ sourced by Eq.~\eqref{eq:nuc_A_j} is a solution to the Yang-Mills equations
	\begin{align}
		\DD_\mu \FF^{\mu\nu}(x^+, x^-, \xperp) = \JJ^\nu_A(x^+, \xperp), \label{eq:nuc_A_ym}
	\end{align}
	with the gauge covariant derivative
	\begin{align}
		\DD_\mu \FF^{\mu\nu} &= \p_\mu \FF^{\mu\nu} - i g \left[ \AA_\mu , \FF^{\mu\nu}\right],
	\end{align}
	and the non-Abelian field strength tensor given by
	\begin{align}
		\FF^{\mu\nu} = \p^\mu \AA^\nu - \p^\nu \AA^\mu - i g \left[ \AA^\mu, \AA^\nu \right].
	\end{align}
	In covariant gauge, $\p_\mu \AA^\mu = 0$, and using appropriate boundary conditions in the asymptotic past $x^0 \rightarrow -\infty$, Eq.~\eqref{eq:nuc_A_ym} is  solved by
	\begin{align}
		\AA^-_{A}(x^+, \xperp) &= \phi_A(x^+, \xperp) = - (\nabla^2_\perp)^{-1} \rho_A(x^+, \xperp), \label{eq:nuc_A_gf}
	\end{align}
	with all other components of $\AA^\mu$ vanishing. The current and color field in Eqs.~\eqref{eq:nuc_A_j} and \eqref{eq:nuc_A_gf} solve the gauge covariant continuity equation
	\begin{align}
		\DD_\mu \JJ^\mu(x) = 0.
	\end{align}
	Similarly, we can consider a nucleus moving along $x^-$ (denoted as ``$B$'') with the analogous current and color field
	\begin{align}
		\JJ^\mu_B(x^-, \xperp) &= \delta^\mu_+ \rho^a_B(x^-, \xperp) t^a, \label{eq:nuc_B_j} \\
		\AA^+_{B}(x^-, \xperp) &= \phi_B(x^-, \xperp) = - (\nabla^2_\perp)^{-1} \rho_B(x^-, \xperp). \label{eq:nuc_B_gf} 
	\end{align}
	
	In order to describe a collision of two nuclei using the Yang-Mills equations we need to solve the collision problem given by
	\begin{align}
		D_\mu F^{\mu\nu}(x) &= J^\mu_A(x) + J^\mu_B(x), \label{eq:full_ym} \\
		D_\mu J^\mu_A(x) &= 0, \label{eq:full_ym_ja} \\
		D_\mu J^\mu_B(x) &= 0, \label{eq:full_ym_jb}
	\end{align}
	with initial conditions specified in the asymptotic past
	\begin{align}
		\lim_{x^0 \rightarrow -\infty} A^\mu(x) &=\AA^\mu_A(x^+, \mathbf x_\perp) + \AA^\mu_B(x^-, \mathbf x_\perp)= \delta^\mu_- \, \phi_A(x^+, \xperp) + \delta^\mu_+ \, \phi_B(x^-, \xperp), \\
		\lim_{x^0 \rightarrow -\infty} J_A^\mu(x) &=\JJ^\mu_A(x^+, \xperp)= \delta^\mu_- \rho_A(x^+, \xperp),  \\
		\lim_{x^0 \rightarrow -\infty} J_B^\mu(x) &=\JJ^\mu_B(x^-, \xperp)= \delta^\mu_+ \rho_B(x^-, \xperp),
	\end{align}
	where we use calligraphic letters $(\mathcal{A},\mathcal{J})$  for the single nuclei solutions, while non-calligraphic letters $(A,J)$ denote the solution for the collision problem (cf.~discussion around Eq.~(\ref{eq:Splitting})).
	In general, there are no closed form solutions for Eq.~\eqref{eq:full_ym}. However, in the ultrarelativistic limit where nuclei become infinitesimally thin, i.e.
	\begin{align}
		\rho_{A/B}(x^\pm, \xperp) = \delta(x^\pm) \bar \rho_{A/B}(\xperp), \label{eq:bi_approx}
	\end{align}
	the solution to Eq.~\eqref{eq:full_ym} becomes invariant under boosts along $z = x^3$ and a partial analytic solution is feasible.  In this case one finds an analytic solution at the boundary of the future light cone ($x^+ > 0$ with $x^- = 0$ or $x^- > 0$ with $x^+ = 0$)~\cite{Kovner:1995ja}, which provides the initial conditions for the subsequent evolution of the Glasma inside the future light cone. The Glasma initial (or boundary) conditions are most conveniently expressed in terms of proper time $\tau = \sqrt{2 x^+ x^-}$ and space-time rapidity $\eta = \ln(x^+ / x^-) / 2$ coordinates  adapting the Fock-Schwinger $(A^\tau = 0)$ gauge. At the boundary of the future light cone, the gauge fields are then given by~\cite{Kovner:1995ja}
	\begin{align}
		A^i(\tau = 0^+, \xperp) &= \alpha^i_A(\xperp) + \alpha^i_B(\xperp), \label{eq:bi_ics_i} \\
		A^\eta(\tau = 0^+, \xperp) &= \frac{ig}{2} \left[\alpha^i_A(\xperp), \alpha^i_B(\xperp)\right], \label{eq:bi_ics_eta}
	\end{align}
	where the color fields $\alpha^i_{A/B}$ are given by
	\begin{align}
		\alpha^i_{A/B}(\xperp) &= \frac{-i}{g} V_{A/B}(\xperp) \p^i V^\dg_{A/B}(\xperp),
	\end{align}
	with the lightlike Wilson lines
	\begin{align}
		V^\dg_A(\xperp) &= \lim_{x^+ \rightarrow \infty} \mathcal{P} \exp \bigg( ig \intop_{-\infty}^{x^+} dx'^+ \phi_A(x'^+, \xperp) \bigg), \\
		V^\dg_B(\xperp) &= \lim_{x^- \rightarrow \infty} \mathcal{P} \exp \bigg( ig \intop_{-\infty}^{x^-} dx'^- \phi_B(x'^-, \xperp) \bigg).
	\end{align}
	Equations \eqref{eq:bi_ics_i} and \eqref{eq:bi_ics_eta} serve as initial conditions at $\tau = 0^+$ for the Yang-Mills equations in the future light cone for $\tau > 0$, which are typically solved numerically on a lattice \cite{Lappi:2003bi,Romatschke:2006nk,Fukushima:2011nq,Schenke:2012wb} or through other approximations such as a Taylor expansion in proper time $\tau$ \cite{Fries:2005yc,Fukushima:2007yk,Fujii:2008km,Chen:2015wia,Carrington:2020ssh}. By construction, solutions in the boost-invariant high-energy limit do not depend on the space-time rapidity $\eta$. 
	
	To model collisions at finite energies, it is necessary to go beyond the boost-invariant approximation given by Eq.~\eqref{eq:bi_approx} and allow for a more general structure of the color charge densities $\rho_{A/B}(x^\pm, \xperp)$ which exhibit a non-trivial dependence on the light cone coordinates. Numerical solution methods using 3+1 dimensional real-time lattice simulations have been previously developed by the authors either based on the colored particle-in-cell method (CPIC) \cite{Gelfand:2016yho, Ipp:2017lho, Ipp:2018hai, Ipp:2020igo} or based on dynamically updated color currents \cite{Schlichting:2020wrv}. Such methods rely on a direct solution of Eqs.~\eqref{eq:full_ym} -- \eqref{eq:full_ym_jb} in the $(t,z)$ coordinate frame and allow to numerically determine fully non-perturbative solutions for the Glasma in 3+1 dimensions. However, due to the large lattice sizes required for stable and accurate simulations, exploring realistic heavy ion collision scenarios using these methods is highly computationally demanding.

	In this work we explore a different approach based on the weak field approximation to obtain semi-analytical approximations to Eqs.~\eqref{eq:full_ym} -- \eqref{eq:full_ym_jb} beyond the boost-invariant limit. The weak field (or dilute) approximation is a perturbative expansion in the color charge densities $\rho_{A/B}(x^\pm, \xperp)$ of the projectile and target. It relies on an explicit split of the gauge field $A^\mu$ into background fields $\AA^\mu$ and perturbations $a^\mu$ 
	\begin{align}
		\label{eq:Splitting}
		A^\mu(x) &= \AA^\mu_{A}(x)+\AA^\mu_{B}(x) + a^\mu(x), \nonumber \\
		J^\mu(x) &= \JJ^\mu_{A}(x)+\JJ^\mu_{B}(x) + j^\mu(x),
	\end{align}
	where the background fields $\AA^\mu_{A/B}$ and background currents $\JJ^\mu_{A/B}$ are given by the single nuclei solutions Eqs.~\eqref{eq:nuc_A_j}, \eqref{eq:nuc_A_gf} and Eqs.~\eqref{eq:nuc_B_j}, \eqref{eq:nuc_B_gf}. Evidently, the details of this perturbative expansion depend on the choice of gauge and we adapt covariant gauge
	\begin{equation}
		\p_\mu A^\mu = \p_\mu \AA^\mu_A + \p_\mu \AA^\mu_B + \p_\mu a^\mu   = 0
	\end{equation}
	throughout this paper. Since the background fields in Eqs.~(\mbox{\ref{eq:nuc_A_gf}) and (\ref{eq:nuc_B_gf}}) readily satisfy $\p_\mu \AA^\mu_A = \p_\mu \AA^\mu_B =0$ this implies $\p_\mu a^\mu = 0$. We further note, that this gauge choice simplifies our calculation, because
	the covariant gauge background fields $\AA^\mu_{A/B} = \OO(\rho_{A/B})$ are linear functionals of the background currents $\JJ^\mu_{A/B}  = \OO(\rho_{A/B})$. 
	These solutions are non-perturbative solutions of the single nucleus problem in the sense that they exactly solve the non-linear Yang-Mills equations (to all orders $\OO(\rho_{A}^{n})$ and $\OO(\rho_{B}^{n})$) before the collision takes place, i.e.~before the currents of the left- and right-moving nuclei start to overlap with each other. Conversely, the perturbative fields $a^\mu$ and $j^\mu$  capture all higher order corrections $\OO(\rho_{A}^n\rho_{B}^m)$ with both $n,m \geq 1$, induced by the interaction of the colliding nuclei. Furthermore, as shown in Eq.~\mbox{\eqref{eq:cov_pert_ym}}, the perturbative field equations for $a^\mu$ also simplify in covariant gauge. While the color field $a^\mu$ describes the (dilute) Glasma itself, the currents $j^\mu$ represent perturbations of the color currents $\JJ^\mu_A$ and $\JJ^\mu_B$ of nuclei $A$ and $B$, due to non-Abelian color rotation. Expanding to the first non-trivial order $\mathcal{O}(\rho_{A}\rho_{B})$ in the color charge densities, the background field equations remain of the same form 
	\begin{align}
		\DD^{A/B}_\mu \FF^{\mu\nu}_{A/B}(x) &= \JJ^\nu_{A/B}(x), \\
		\DD^{A/B}_\mu \JJ^\mu_{A/B}(x) &= 0,
	\end{align}
	absorbing all terms of $\mathcal{O}(\rho_{A}^{n}\rho_{B}^{0})$ and respectively $\mathcal{O}(\rho_{A}^{0}\rho_{B}^{n})$, while the perturbative field equations that account for the interaction of the nuclei read 
	\begin{align}
		\p_\mu f^{\mu\nu}(x) -ig[\AA_{\mu}^{A}(x), \FF^{\mu\nu}_{B}(x)]  -ig[\AA_{\mu}^{B}(x), \FF^{\mu\nu}_{A}(x)]= j^\nu(x), \label{eq:pert_ym} \\
		\p_\mu j^\mu(x) = + i g \, \left[ \AA_\mu^{A}(x), \JJ^\mu_{B}(x) \right]+ i g \, \left[ \AA_\mu^{B}(x), \JJ^\mu_{A}(x) \right], \label{eq:pert_cont}
	\end{align}
	with
	\begin{align}
		f^{\mu\nu}(x) = \p^\mu a^\nu(x) - \p^\nu a^\mu(x) -ig \left[ \AA^\mu_{A}(x), \AA^\nu_{B}(x)\right]-ig \left[ \AA^\mu_{B}(x), \AA^\nu_{A}(x)\right].
	\end{align}
	Since the perturbations represent the Glasma created from the collision of the two nuclei, we assume that both perturbative fields and currents vanish in the asymptotic past
	\begin{align}
		\lim_{x^0 \rightarrow -\infty} \, a^\mu(x) &= 0, \label{eq:a_asy_past} \\
		\lim_{x^0 \rightarrow -\infty} \, j^\mu(x) &= 0. \label{eq:j_asy_past}
	\end{align}
	Assuming that the color charges of the colliding nuclei do not change their trajectories and considering the initial conditions in Eq.~\eqref{eq:j_asy_past}, the solution to Eq.~\eqref{eq:pert_cont} is straightforward:
	\begin{align}
		j^+(x^+, x^-, \xperp) &= ig \intop_{-\infty}^{x^+} dz^+ \left[ \phi_A(z^+, \xperp), \rho_B(x^-, \xperp) \right], \\
		j^-(x^+, x^-, \xperp) &= ig \intop_{-\infty}^{x^-} dz^- \left[ \phi_B(z^-, \xperp), \rho_A(x^+, \xperp) \right].
	\end{align}
	In covariant gauge, $\p_\mu a^\mu = 0$, Eq.~\eqref{eq:pert_ym} simplifies to 
	\begin{align}
		\p^2 a^\mu(x) = S^\mu(x), \label{eq:cov_pert_ym}
	\end{align}
	where $S^\mu(x)$ are $a^\mu$-independent source terms given by
	\begin{align}
		S^+(x^+, x^-, \xperp) &= + ig \big( \p_- \left[ \phi_A(x^+, \xperp), \phi_B(x^-, \xperp) \right] +  \intop_+ \left[ \phi_A(x^+, \xperp), \rho_B(x^-, \xperp) \right] \big), \label{eq:S_p} \\
		S^-(x^+, x^-, \xperp) &= - ig \big( \p_+ \left[ \phi_A(x^+, \xperp), \phi_B(x^-, \xperp) \right] +  \intop_- \left[ \rho_A(x^+, \xperp), \phi_B(x^-, \xperp) \right] \big), \label{eq:S_m} \\
		S^i(x^+, x^-, \xperp) &= - ig \big( \left[ \phi_A(x^+, \xperp), \p^i \phi_B(x^-, \xperp) \right] - \left[ \p^i \phi_A(x^+, \xperp), \phi_B(x^-, \xperp) \right] \big). \label{eq:S_i}
	\end{align}
	Here, we have introduced a slightly unusual but very useful shorthand
	\begin{align}
		\intop_\pm f(x^\pm) &\equiv \intop^{x^\pm}_{-\infty} dz^\pm f(z^\pm).
	\end{align}
	Due to the choice of covariant gauge, we can independently solve for the four independent components of $a^\mu$ in
	Eq.~\eqref{eq:cov_pert_ym}. Analyzing the $x^\pm$ dependence of the source terms in Eqs.~\eqref{eq:S_p} -- \eqref{eq:S_i}, we find that $S^\pm(x)$ are only non-zero along the boundaries of the future light cone, whereas $S^i(x)$ only has support in the vicinity of the collision center $x^+ = x^- = 0$. Based on the initial conditions in Eq.~\eqref{eq:a_asy_past}, we can then formally solve the field equations in Eq.~\eqref{eq:cov_pert_ym} as
	\begin{align} \label{eq:amu_conv}
		a^\mu(x) = \intop_y \Gret(x-y) S^\mu(y),
	\end{align}
	where $\Gret(z)$ denotes the retarded propagator
	\begin{align}
		\Gret(z) = - \frac{1}{2\pi} \Theta(z^0) \delta(z^\mu z_\mu),
	\end{align}
	ensuring causality in compliance with the initial conditions.
	
	\subsection{Gauge field solutions in the future light cone}
	We now focus on carrying out the integration in Eq.~\eqref{eq:amu_conv} as far as possible to find simple expressions for the gauge field $a^\mu$ in terms of color potentials $\phi^a_{A/B}$ of the colliding nuclei. 
	We start by noting that the source terms in Eqs.~\eqref{eq:S_p} -- \eqref{eq:S_i} can be stated in a more unified way by performing a partial Fourier transform over transverse coordinates. We use
	\begin{align}
		\rho^a_A(x^+, \mathbf x_\perp) = \intop_{\mathbf p_\perp} \, \tilde \rho^a_A(x^+, \mathbf p_\perp) e^{- i \mathbf p_\perp \cdot  \mathbf x_\perp},
	\end{align}
	with $\intop_{\mathbf p_\perp} = \intop \frac{d^2  \mathbf p_\perp}{(2\pi)^2}$ and rewrite the charge densities in terms of potentials via ${\tilde \rho^a(x^+, \mathbf p_\perp) = \mathbf p^2_\perp \tilde{\phi}^a(x^+, \mathbf p_\perp)}$. Here, we use $\mbf p_\perp \cdot \mbf x_\perp = p_i x_i$ and $\mbf p_\perp^2 = p_i p_i$. The source terms are then simply 
	\begin{align}
		S^+(x) &=  \intop_{\mathbf p_\perp} \intop_{\mathbf q_\perp}  \Big( + \p_- + \mathbf q^2_\perp \intop_{+} \Big)  S_d(x^+, x^-, \mathbf  p_\perp, \mathbf  q_\perp) e^{-i(\mathbf p+ \mathbf q)_\perp \cdot \mathbf  x_\perp}, \label{eq:S_p_dummy} \\ 
		S^-(x) &=  \intop_{\mathbf p_\perp} \intop_{\mathbf q_\perp}  \Big( -\p_+ - \mathbf  p^2_\perp \intop_{-} \Big)  S_d(x^+, x^-, \mathbf  p_\perp, \mathbf  q_\perp)  e^{-i(\mathbf  p+\mathbf  q)_\perp \cdot \mathbf  x_\perp}, \label{eq:S_m_dummy} \\
		S^i(x) &=  \intop_{\mathbf p_\perp} \intop_{\mathbf q_\perp}  i \big(\mathbf  p^i_\perp -  \mathbf  q^i_\perp \big)  S_d(x^+, x^-, \mathbf p_\perp, \mathbf q_\perp)   e^{-i(\mathbf  p+\mathbf  q)_\perp \cdot \mathbf x_\perp}\label{eq:S_i_dummy},
	\end{align}
	where we defined the auxiliary source term $S_d$ as
	\begin{align}
		S_d(x^+, x^-, \mathbf  p_\perp, \mathbf  q_\perp) = - g f^{abc} t^c \tilde{\phi}^a_A(x^+, \mathbf  p_\perp) \tilde{\phi}^b_B(x^-, \mathbf  q_\perp) \label{eq:S_dummy}.
	\end{align}
	Our strategy for performing integrations in Eq.~\eqref{eq:amu_conv} is most easily demonstrated using the transverse gauge field $a^i$, where it is easy to see that the transverse integration only acts on the phase factor:
	\begin{align} \label{eq:transverse_field_intermediate}
		a^i(x) &= \intop_y \Gret(x - y) S^i(y) \nonumber \\
		&= \intop_{-\infty}^{+\infty} dy^+ \intop_{-\infty}^{+\infty} dy^- \int d^2 \mathbf y_\perp \, \Gret(x^+ \! - \! y^+, x^- \! - \!  y^-, \mathbf x_\perp \! - \!  \mathbf y_\perp) S^i(y^+, y^-, \mathbf y_\perp) \nonumber \\
		&= \intop_{-\infty}^{+\infty} dy^+ \intop_{-\infty}^{+\infty} dy^- \intop_{\mathbf p_\perp} \intop_{\mathbf q_\perp}  i (\mathbf p^i_\perp \! - \! \mathbf q^i_\perp)
		S_d(y^+, y^-, \mathbf p_\perp, \mathbf q_\perp)
		\intop d^2 \mbf y_\perp \Gret(x \! - \! y) e^{-i(\mathbf p+\mathbf q)_\perp \cdot \mathbf y_\perp}.
	\end{align}
	By performing a change of variables $z^\mu = x^\mu - y^\mu$ we can then solve the integral over $\mathbf y_\perp$ as
	\begin{align}
		&\intop d^2 \mathbf y_\perp \Gret(x\! - \! y) e^{-i(\mbf p+\mbf q)_\perp \cdot \mbf y_\perp} = \nonumber \\
		&\qquad -e^{-i(\mbf p+\mbf q)_\perp \cdot \mbf x_\perp}  \intop^\infty_0 d|\mathbf z_{\perp}| |\mathbf z_{\perp}| \intop^{2\pi}_0 d\phi \frac{1}{2\pi} \Theta(z^{0}) \delta(2 z^+ z^- \! - \! |\mathbf z_{\perp}|^2) e^{+i |p+q| |\mathbf z_{\perp}| \cos \phi},
	\end{align}
	where $\cos\phi=\frac{\mathbf z_{\perp} \cdot (\mathbf p_{\perp}+\mathbf q_{\perp})}{|\mathbf z_{\perp}|| \mathbf p_{\perp}+\mathbf q_{\perp}|}$ is the azimuthal angle between $\mathbf z_{\perp}$ and $\mathbf p_{\perp} \! + \! \mathbf q_{\perp}$ and we denote
	$|p+q| = |(\mbf p+\mbf q)_\perp|$. Evaluating the $\phi$ and $|\mathbf z_{\perp}|$ integrals, we  then obtain
	\begin{align}
		\intop d^2 \mathbf y_\perp \Gret(x-y) e^{-i(\mbf p+\mbf q)_\perp \cdot \mbf y_\perp}&= - e^{-i(\mbf p+ \mbf q)_\perp \cdot \mbf x_\perp} \Theta(z^0) \intop^\infty_0 d|\mathbf z_{\perp}| |\mathbf z_{\perp}| \, \delta(2 z^+ z^- \! - \! |\mathbf z_{\perp}|^2) J_0(|p \! + \! q| |\mathbf z_{\perp}|)\nonumber \\
		&= - \frac{1}{2} \Theta(z^0) \Theta(\tau_z)  J_0(|p \! + \! q| \tau_z) e^{-i(\mbf p+\mbf q)_\perp \cdot \mbf x_\perp} ,
	\end{align}
	where $\tau_z = \sqrt{2 z^+ z^-}$ and the two Heaviside functions imply that this term  only contributes in the future light cone.  By inserting this result into Eq.~\eqref{eq:transverse_field_intermediate}, we obtain
	\begin{align}
		a^i(x) = - \frac{1}{2}  \intop_{p_\perp} \intop_{q_\perp}  i (\mbf p^i_\perp \! - \! \mbf q^i_\perp) \intop^\infty_0 dz^+ \intop^\infty_0 dz^- 
		S_d(x^+ \! - \!  z^+, x^- \! - \!  z^-, \mbf p_\perp,  \mbf q_\perp)J_0(|p \! + \! q| \tau_z) e^{-i(\mbf p \! + \mbf q)_\perp \cdot \mbf x_\perp},
	\end{align}
	which is more compactly written as
	\begin{align}
		a^i(x) &= \intop_{p_\perp} \intop_{q_\perp}  i (\mbf p^i_\perp \! - \! \mbf q^i_\perp) a_d(x^+, x^-, \mbf p_\perp, \mbf q_\perp) e^{-i( \mbf p+ \mbf q)_\perp \cdot \mbf x_\perp}, \label{eq:ai}
	\end{align}
	with the auxiliary field $a_d$ given by
	\begin{align}\label{Eq:DummyField}
		a_d(x^+, x^-, \mbf p_\perp,  \mbf q_\perp) &= \frac{g}{2} f_{abc} t^c \intop^\infty_0 dz^+ \intop^\infty_0 dz^- 
		\tilde{\phi}^a_A(x^+ \! - \! z^+, \mbf p_\perp) \tilde{\phi}^b_B(x^- \! - \! z^-, \mbf q_\perp) J_0(|p \! + \! q| \tau_z).
	\end{align}
	Carrying out the same steps for $a^+$ and $a^-$, we find analogous expressions
	\begin{align}
		a^+(x) &= \intop_{\mathbf p_\perp} \intop_{\mathbf q_\perp}   \Big( +\p_- + \mbf q^2_\perp \intop_{+} \Big)
		a_d(x^+, x^-, \mbf p_\perp, \mbf q_\perp) e^{-i( \mbf p+ \mbf q)_\perp \cdot \mbf x_\perp},\label{eq:ap} \\
		a^-(x) &= \intop_{\mathbf p_\perp} \intop_{\mathbf q_\perp}   \Big( -\p_+- \mbf p^2_\perp \intop_{-} \Big)
		a_d(x^+, x^-, \mbf  p_\perp, \mbf q_\perp) e^{-i( \mbf p+ \mbf q)_\perp \cdot \mbf x_\perp} \label{eq:am},
	\end{align}
	and it is straightforward to check that Eqs.~\eqref{eq:ai} -- \eqref{eq:am} satisfy the gauge condition
	\begin{align}
		\p_\mu a^ \mu = \p_+ a^+ + \p_- a^- + \p_i a^i = 0.
	\end{align}
	
	We can further simplify the expressions for $a^+$ and $a^-$ by explicitly computing the derivatives and integrals with respect to $x^\pm$ in Eqs.~\eqref{eq:ap} and \eqref{eq:am}. Starting with the derivative term in Eq.~\eqref{eq:ap}, we use integration by parts to find
	\begin{align}
		\p^{(x)}_- a_d
		&= \frac{g}{2} f_{abc} t^c \intop^\infty_0 dz^+ \intop^\infty_0 dz^- \tilde \phi^a_A(x^+ \! - \! z^+, \mbf p_\perp) \tilde \phi^b_B(x^- \! - \! z^-, \mbf q_\perp) \p^{(z)}_- J_0(|p \! + \! q| \tau_z) \nonumber \\
		&\quad + \frac{g}{2} f_{abc} t^c\intop^\infty_0 dz^+ \tilde \phi^a_A(x^+ \! - \! z^+, \mbf p_\perp) \tilde \phi^b_B(x^-, \mbf q_\perp).
		\label{eq:WurstSalat}
	\end{align}
	The second line in the above expression is the boundary term for $z^- \rightarrow 0$, which generally does not vanish in contrast to the $z^- \rightarrow \infty$ boundary. However, it is proportional to the color potential $ \tilde \phi^b_B(x^-, \mathbf q_\perp)$, which vanishes inside the future light cone. If we are only interested in far field solutions, we can safely ignore this term. By use of the following relations
	\begin{align}
		\p_\pm \tau_z &= z^\mp / \tau_z, \\
		\p_\pm J_0(|p+q| \tau_z) &= - J_1(|p+q| \tau_z) |p+q| \frac{z^\mp}{\tau_z}, 
	\end{align}
	we then find
	\begin{align}
		\p^{(x)}_- a_d
		&\simeq - \frac{g}{2} f_{abc} t^c \intop^\infty_0 dz^+ \intop^\infty_0 dz^- \tilde \phi^a_A(x^+ \! - \! z^+, \mbf p_\perp) \tilde \phi^b_B(x^- \! - \! z^-, \mbf q_\perp) |p \! + \! q| \frac{z^+}{\tau_z} J_1(|p \! + \! q| \tau_z), \label{eq:ap_der_term}
	\end{align}
	where we use $\simeq$ to denote that, due to the fact that we have ignored the boundary terms in the second line of Eq.~(\ref{eq:WurstSalat}), this expression is only strictly valid inside the future light cone.
	The term involving an integration in Eq.~\eqref{eq:ap} is given by
	\begin{align}
		& \int_+ a_d ( x^+, x^-, \mbf p_\perp, \mbf q_\perp) = \intop_{-\infty}^{x^+} d \tilde x^+ a_d(\tilde x^+, x^-, \mbf p_\perp, \mbf q_\perp) \nonumber \\
		&= \frac{g}{2} f_{abc} t^c \intop_{-\infty}^{x^+} d \tilde x^+ \intop_0^{+\infty} d \tilde z^+ \intop_0^{+\infty} dz^- \tilde \phi^a_A(\tilde x^+ \! - \! \tilde z^+, \mbf p_\perp) \tilde \phi^b_B(x^- \! - \! z^-, \mbf q_\perp) J_0(|p \! + \! q| \sqrt{2 \tilde z^+ z^-}) \nonumber \\
		&= \frac{g}{2} f_{abc} t^c \intop_0^{+\infty} dz^+ \intop_0^{+\infty} dz^- \tilde \phi^a_A( x^+ \! - \! z^+, \mbf p_\perp) \tilde \phi^b_B(x^- \!  - \! z^-, \mbf q_\perp) \intop_0^{z^+} d \tilde z^+ \, J_0(|p\! + \! q| \sqrt{2 \tilde z^+ z^-}), \label{eq:ap_int_term}
	\end{align}
	where we have performed a change of variables from $\tilde{x}^{+}$ to $z^{+}=x^{+}-\tilde{x}^{+}+\tilde{z}^{+}$ to isolate the integration over $\tilde z^+$ in the last line.\footnote{It is instructive to express the terms in the second line of Eq.~\eqref{eq:ap_int_term} with two placeholder functions $f(\tilde x^+ - \tilde z^+)$ and $g(\tilde z^+)$ and re-formulate the integral bounds in terms of Heaviside functions as ${\intop_{0}^{+\infty} d \tilde z^+ \intop_{-\infty}^{x^+} d \tilde x^+ f(\tilde x^+ - \tilde z^+) g( \tilde z^+) =} \intop_{-\infty}^{+\infty} d \tilde z^+ \intop_{-\infty}^{+\infty} d\tilde x^+ f(\tilde x^+ - \tilde z^+) g( \tilde z^+) \Theta( \tilde z^+) \Theta(x^+ - \tilde x^+)$. By performing a change of variables, expressing  $z^{+}=x^{+}-\tilde x^{+} +\tilde{z}^{+}$ such that $f(\tilde x^+ - \tilde z^+)=f(x^+ -z^+)$, one then finds that the Heaviside functions constrain the integration domain to $0 < \tilde z^+ < z^+$.} The integral over the Bessel function is given by
	\begin{align}
		\intop_0^{z^+} d \tilde z^+ \, J_0(|p+q| \sqrt{2 \tilde z^+ z^-}) = \frac{1}{|p+q|} \frac{\tau_z}{z^-} J_1( |p+q| \tau_z),
	\end{align}
	which leads to
	\begin{align}
		& \intop_{-\infty}^{x^+} d \tilde x^+ a_d(\tilde x^+, x^-, \mbf p_\perp, \mbf q_\perp) \nonumber \\
		&= \frac{g}{2} f_{abc} t^c \intop_0^{+\infty} dz^+ \intop_0^{+\infty} dz^- \tilde \phi^a_A(x^+ \! - \! z^+, \mbf p_\perp) \tilde \phi^b_B(x^- \! - \! z^-, \mbf q_\perp) \frac{1}{|p\! + \! q|} \frac{\tau_z}{z^-} J_1( |p \! + \! q| \tau_z). \label{eq:ap_int_term2}
	\end{align}
	Inserting the results in Eqs.~\eqref{eq:ap_der_term} and \eqref{eq:ap_int_term2} into the expression Eq.~(\ref{eq:ap}), we finally obtain
	\vspace{0.5em}
	\begin{tcolorbox}\vspace{-1em}\begin{align}\label{eq:ap_final} 
			a^+(x) & \simeq  \frac{g}{2} f_{abc} t^c  \intop_{\mathbf p_\perp,\mathbf q_\perp}  \intop_0^{+\infty}   dz^+   \intop_0^{+\infty} dz^- \tilde \phi^a_A( x^+   -  z^+ \!, \mbf p_\perp) \tilde \phi^b_B(x^-  -  z^- \! , \mbf q_\perp) \quad \nonumber \\
			& \qquad\times \frac{
				\big( - \! (\mbf p_\perp \! + \! \mbf q_\perp)^2 + 2 \mbf q_\perp^2 \big) z^+
			}{|\mathbf p_{\perp} \! + \! \mathbf q_{\perp} | \tau_z}
			J_1( |\mathbf p_{\perp} \! + \! \mathbf q_{\perp} | \tau_z) e^{-i(\mbf p+ \mbf q)_\perp \cdot \mbf x_\perp}.
	\end{align}\end{tcolorbox}\noindent
	By repeating the same analogous steps for the calculation of $a^{-}$, we obtain
	\vspace{0.5em}
	\begin{tcolorbox}\vspace{-1em}\begin{align} 
			\label{eq:am_final} 
			a^-(x) &\simeq \frac{g}{2} f_{abc} t^c \intop_{\mathbf p_\perp ,\mathbf q_\perp}\intop_0^{+\infty} dz^+ \intop_0^{+\infty} dz^- \tilde \phi^a_A( x^+ \! - \! z^+,\mbf p_\perp) \tilde \phi^b_B(x^- \! - \! z^-, \mbf q_\perp) \quad  \nonumber \\ 
			& \qquad\times \frac{\big(+(\mbf p_\perp \! + \! \mbf q_\perp)^2 - 2 \mbf p_\perp^2\big) z^-}{ |\mathbf p_{\perp} \! + \! \mathbf q_{\perp} | \tau_z }
			J_1( |\mathbf p_{\perp} \! + \! \mathbf q_{\perp} | \tau_z) e^{-i(\mbf p+ \mbf q)_\perp \cdot \mbf x_\perp}, 
		\end{align}
	\end{tcolorbox}\noindent
	which together with the transverse components of the gauge fields in Eqs.~\eqref{eq:ai} and \eqref{Eq:DummyField}
	\begin{tcolorbox}\vspace{-1em}\begin{align}
			\label{eq:ai_final} 
			a^i(x) &= \frac{g}{2} f_{abc} t^c \intop_{\mathbf p_\perp ,\mathbf q_\perp}\intop_0^{+\infty} dz^+ \intop_0^{+\infty} dz^- \tilde \phi^a_A( x^+ \! - \! z^+,\mbf p_\perp) \tilde \phi^b_B(x^- \! - \! z^-, \mbf q_\perp) \quad  \nonumber \\ 
			& \qquad \times i(\mathbf p_\perp^i- \mathbf q_\perp^i)~J_0( |\mathbf p_{\perp} \! + \! \mathbf q_{\perp} | \tau_z) e^{-i(\mbf p+ \mbf q)_\perp \cdot \mbf x_\perp}, 
	\end{align}\end{tcolorbox}\noindent
	provide our final expressions for the Glasma fields in the future light cone.
	\section{Nuclear model and transverse pressure}\label{AnalyticSect}
	Based on the previous analytical calculation, the longitudinal structure of the Glasma at late times can be obtained by considering a specific model for the color charge distribution inside a nucleus. Within this study, we consider a simple McLerran-Venugopalan (MV) type model \cite{McLerran:1993ka,McLerran:1993ni} of a transversally homogeneous nucleus, where fluctuations of the color charge density are given by 
	\begin{align}
		\ev{\rho^a_{A/B}(x^\pm, \mathbf{x_\perp}) \rho^b_{A/B}(x'^\pm, \mathbf x'_\perp) } = g^2 \mu^2_{A/B}\delta^{ab} T_R(\frac{x^\pm + x'^\pm}{2}) U_\xi(x^\pm - x'^\pm)\GMV(\mathbf x_\perp- \mathbf x'_\perp). \label{eq:corr_wf}
	\end{align}
	The constant $g^2\mu_{A/B}$ denotes the color density per unit transverse area which is related to saturation momentum $Q_s$, while the function $\GMV$ characterizes the transverse correlation of color charges inside the nucleus. Similarly, the functions $T_R$ and $U_\xi$
	describe the longitudinal profile and correlations of color charges, and are taken as normalized Gaussians with widths $R$ and $\xi$ identified as the Lorentz contracted size of the nucleus and longitudinal correlation length respectively.  In order to enforce color neutrality on average, the one-point function is assumed to be zero.
	
	Using this model, we can use our previous results to investigate a wide range of observables. In particular, we are interested in the various components of the energy-momentum tensor given by
	\begin{align}
		&T^{\mu\nu} = - F^{a,\mu\rho} F^{a,\nu}{}_{\rho} + \frac{1}{4} g^{\mu\nu} F^{a,\rho\sigma} F^a_{\rho\sigma} \notag\\
		&= 2\, \mathrm{Tr} \Big[-\underbrace{\FF^{\mu\rho}\FF^\nu{}_\rho+\frac{1}{4}g^{\mu\nu}\FF^{\rho\sigma}\FF_{\rho\sigma}}_{\rm{Background}}-\underbrace{\big(\FF^{\mu\rho}f^\nu{}_\rho+f^{\mu\rho}\FF^\nu{}_\rho\big)+\frac{1}{2}g^{\mu\nu}f^{\rho\sigma}\FF_{\rho\sigma}}_{\rm{Mixed~term}}-\underbrace{f^{\mu\rho}f^\nu{}_\rho+\frac{1}{4}g^{\mu\nu}f^{\rho\sigma}f_{\rho\sigma}}_{\rm{Perturbative}} \big],
	\end{align}
	where we have made the split into background, mixed and perturbative terms explicit.
	For the purposes of this paper, we only consider the perturbative part of the energy-momentum tensor. We have derived the perturbative field $a^\mu$ up to quadratic order, $\OO(\rho_A \rho_B)$, which yields the perturbative energy-momentum tensor up to quartic order, $\OO(\rho_A^2 \rho_B^2)$. In principle, the mixed terms could also contain quartic contributions, however, the background field strengths $\FF^{\mu\nu}$ are only non-zero along the light cone. Consequently, the mixed terms vanish inside the future light cone to all orders. Since we are only interested in the Glasma, we can safely ignore the mixed terms. In the following we focus on the transverse pressure which is solely generated during the collision and hence has no contribution from the background and the mixed part outside the space-time region where the colliding nuclei overlap.
	The transverse pressure is given by 
	\begin{align}
		p_T = \frac{T^{ii}}{2} = \varepsilon_{E,L} + \varepsilon_{B,L},
	\end{align}
	where $\varepsilon_{E,L}$ and $\varepsilon_{B,L}$ are the contributions from the longitudinal electric and longitudinal magnetic field given by
	\begin{align}
		\varepsilon_{E,L}=\ev{\tr f^{+-}f^{+-}},\\ 
		\varepsilon_{B,L}=\frac{1}{2} \ev{\tr f^{ij} f_{i j}}.
	\end{align}
	\subsection{Longitudinal magnetic field}
	To get the longitudinal magnetic field, we first calculate the corresponding field strength $f_{ij}$ with Eq.~\eqref{eq:ai}
	\begin{align}
		f_{ij} = \p_i a_j - \p_j a_i =  2\intop_{\mathbf p_\perp, \mathbf q_\perp}(p^i q^j - q^i p^j)  a_d(x^+, x^-, \mathbf p_\perp, \mathbf q_\perp) e^{-i(\mathbf p_\perp+ \mathbf q_\perp) \cdot \mathbf x_\perp}.
	\end{align}
	The square of the above expression contains integrals over four color potentials, arising from the auxiliary fields $a_d$. Since it is quite convenient to solve such integrals in Fourier space, we write the correlation function in Eq.~\eqref{eq:corr_wf} as
	\begin{align}\label{eq:correlator_MS}
		\ev{\tilde \rho^a_{A/B}(x^+, \mathbf p_\perp) \tilde \rho^b_{A/B}(x'^+, \mathbf q_\perp) } = (2\pi)^2  g^2 \mu^2_{A/B}\delta^{ab} T_R(\frac{x^+ \! + \! x'^+}{2}) U_\xi(x^+ \! - \! x'^+)\delta^{(2)}(\mathbf p_\perp \! + \! \mathbf q_\perp)\tilde{\GMV}\left(\frac{\mathbf p_\perp \!- \!\mathbf q_\perp}{2}\right).
	\end{align}
	Exploiting the fact that the nuclear model is diagonal in momentum space, we have
	\begin{align}
		\varepsilon_{B,L} &{}= \frac{1}{2} \ev{\tr f^2_{ij}} \notag \\
		&{}= 4 \intop_{\mathbf p_\perp,\mathbf q_\perp} (\mathbf p_\perp^2 \mathbf q_\perp^2 - (\mathbf p \cdot \mathbf q)^2_\perp) \ev{\tr \, \left[ a_d(x^+, x^-, \mathbf p_\perp, \mathbf q_\perp) a_d(x^+, x^-, -\mathbf p_\perp, -\mathbf q_\perp) \right]} \notag \\
		&{}= \frac{g^2}{2} N_c (N_c^2 - 1) \intop_{\mathbf p_\perp,\mathbf q_\perp}  \intop_{z^\pm} \intop_{\bar{z}^\pm}  (\mathbf p_\perp^2 \mathbf q_\perp^2 - (\mathbf p \cdot \mathbf q)^2_\perp)\ev{\tilde \phi_A(x^+-z^+,\mathbf p_\perp) \tilde \phi_A(x^+-\bar z^+,-\mathbf p_\perp)}  \notag \\
		&{}\qquad\times \ev{\tilde \phi_B(x^--z^-,\mathbf q_\perp) \tilde \phi_B(x^- -\bar z^-,-\mathbf q_\perp)} J_0(|p+q| \tau_z)J_0(|p+q| \tau_{\bar z}) \notag \\
		&{}= \frac{g^2}{2} N_c (N_c^2 - 1) \intop_{\mathbf p_\perp,\mathbf q_\perp}  \intop_{z^\pm} \intop_{\bar{z}^\pm}  J_0(|p+q| \tau_z)J_0(|p+q| \tau_{\bar z}) (\mathbf p_\perp^2 \mathbf q_\perp^2 - (\mathbf p \cdot \mathbf q)^2_\perp) C_A(\mathbf p_\perp)  \notag \\
		&{}\qquad \times C_B(\mathbf q_\perp)T_A(x^+-\frac{z^+ + \bar z^+}{2}) T_B(x^- -\frac{z^- + \bar z^-}{2}) U_A(\bar z^+ - z^+) U_B(\bar z^- - z^-),\label{eq:long_exp_for_B}
	\end{align}
	where we have evaluated the color factors as $f_{abc}f_{abc}=N_c(N_c^2-1)$.
	To obtain the last equality, we have used Eqs.~\eqref{eq:nuc_A_gf} and \eqref{eq:nuc_B_gf}, and replaced the gauge field correlators with our nuclear model such that the overall transverse dependence is characterized by 
	\begin{align}
		C_{A/B}(\mathbf p_\perp) = \frac{g^2 \mu^2_{A/B}\tilde{\GMV}(\mathbf p_\perp)}{\mathbf p_\perp^4},
	\end{align}
	which we take as
	\begin{align}
		C_{A/B}(\mathbf p_\perp) = \frac{g^2 \mu^2}{(\mathbf p^2_\perp + m^2)^2} e^{-\frac{\mathbf p_\perp^2}{\Lambda^2}}
	\end{align}
	for both nuclei $A$ and $B$, where, adopting the same conventions as in \mbox{\cite{Schlichting:2020wrv}}, $m$ and $\Lambda$ regulate the infrared
	and ultraviolet modes respectively. Inspecting the coordinate dependence of Eq.~\eqref{eq:long_exp_for_B}, it is convenient to perform a change of variables from $z^\pm, \bar z^\pm$ to mean and relative coordinates $Z^\pm, \delta z^\pm$
	\begin{align}
		Z^\pm &= \frac{z^\pm + \bar z^\pm}{2}, \\
		\delta z^\pm &= z^\pm - \bar z^\pm.
	\end{align}
	Since $T$ and $U$ are both Gaussian functions, we can change the limits of integration to
	\begin{align}
		\int_0^\infty dz^+dz^-d\bar{z}^+\bar{z}^- = \int_{0}^{\infty} dZ^{+} \int_{-2Z^{+}}^{+2Z^{+}} d \delta z^{+}   \int_{0}^{\infty} dZ^{-} \int_{-2Z^{-}}^{+2Z^{-}} d\delta z^{-}.
	\end{align}
	The resulting expression for the longitudinal magnetic field is then given by
	\begin{align}
		\varepsilon_{B,L} &{}= \frac{g^2}{2} N_c (N_c^2 - 1) \intop_{\mathbf p_\perp,\mathbf q_\perp}  \intop^\infty_0 dZ^+ \intop^\infty_0 dZ^- \intop_{-2Z^+}^{+2Z^+} d\delta z^+ \intop_{-2Z^-}^{+2Z^-} d\delta z^- \big(\mathbf p_\perp^2 \mathbf q_\perp^2 - (\mathbf p \cdot \mathbf q)^2_\perp\big)  \notag \\
		&{} \times C_A(\mathbf p_\perp) C_B(\mathbf q_\perp) T_A(x^+-Z^+) T_B(x^- -Z^-) U_A(\delta z^+) U_B(\delta z^-) \notag\\
		&{}\times J_0(|p+q|\tau_z )J_0(|p+q|\tau_{\bar{z}}), \label{eq:e_BL}
	\end{align}
	with $\tau_z=\sqrt{2(Z^{+}+\delta z^{+}/2) (Z^{-}+\delta z^{-}/2)}
	$ and $\tau_{\bar{z}}=\sqrt{2(Z^{+}-\delta z^{+}/2) (Z^{-}-\delta z^{-}/2)}$.
	
	\subsection{Longitudinal electric field}
	Similarly, in order to calculate the longitudinal electric field, we start again with the associated field strength $f_{+-}$ by using Eqs.~\eqref{eq:ap} and \eqref{eq:am} 
	\begin{align}
		f_{+-} &{}=\partial_+a^+-\partial_-a^-\notag\\
		&{}= \intop_{\mathbf p_\perp,\mathbf q_\perp} \big(2 \p_+ \p_- a_d(x^+, x^-, \mathbf p_\perp, \mathbf q_\perp) + (\mathbf p_\perp^2 + \mathbf q_\perp^2) a_d(x^+, x^-, \mathbf p_\perp, \mathbf q_\perp)\big) e^{-i (\mathbf p+\mathbf q)_\perp \cdot \mathbf x_\perp}.\label{eq:f_pm}
	\end{align}
	Since we have already found the derivative of the auxiliary field $a_d$ in Eq.~\eqref{eq:ap_der_term}, we differentiate it again with respect to $x^+$ to get the first term of $f_{+-}$. We find 
	\begin{align}
		2 \p_+ \p_- a_d &\simeq - g f_{abc} t^c \intop^\infty_0 dz^+ \intop^\infty_0 dz^- \p^{(x)}_+ \tilde \phi^a_A(x^+ \! - \! z^+,\mathbf p_\perp) \tilde \phi^b_B(x^- \! - \! z^-,\mathbf q_\perp) |p+q| \frac{z^+}{\tau_z} J_1(|p+q| \tau_z) \notag \\
		&=- g f_{abc} t^c \intop^\infty_0 dz^+ \intop^\infty_0 dz^-  \tilde \phi^a_A(x^+ \! - \! z^+,\mathbf p_\perp) \tilde \phi^b_B(x^- \! - \! z^-,\mathbf q_\perp) |p+q| \p^{(z)}_+ \left( \frac{z^+}{\tau_z} J_1(|p+q| \tau_z) \right) \notag \\
		&= -\frac{g}{2} f_{abc} t^c  \intop^\infty_0 dz^+ \intop^\infty_0 dz^-  \tilde \phi^a_A(x^+ \! -\! z^+,\mathbf p_\perp) \tilde \phi^b_B(x^- \! - \! z^-,\mathbf q_\perp) |p+q|^2 J_0(|p+q| \tau_z)\label{eq:Q},
	\end{align}
	where the $\simeq$ in the first line denotes the omission of boundary terms that are not relevant within the future light cone, and  we used the Bessel identity
	\begin{align}
		\frac{2}{x} J_1(x) = J_0(x) + J_2(x)
	\end{align} 
	to obtain the final equality. Using the Eq.~\eqref{eq:Q} together with Eq.~\eqref{Eq:DummyField}, we obtain
	\begin{align}
		f_{+-} =  - g f_{abc} t^c \intop_{\mathbf p_\perp,\mathbf q_\perp}  \intop^\infty_0 dz^+ \intop^\infty_0 dz^-  \tilde \phi^a_A(x^+ \! - \! z^+,\mathbf p_\perp) \tilde \phi^b_B(x^- \! - \! z^-,\mathbf q_\perp) (\mathbf p \cdot \mathbf q)_\perp J_0(|p+q| \tau_z) e^{-i (\mathbf p+\mathbf q)_\perp \cdot \mathbf x_\perp}.
	\end{align}
	With this, the expression for the longitudinal electric field takes the following form
	\begin{align}
		\varepsilon_{E,L} &= \ev{\tr \left[ f^2_{+-}\right]} \notag\\
		&= \frac{g^2}{2} N_c (N_c^2 - 1) \intop_{\mathbf p_\perp,\mathbf q_\perp} \intop_{z^\pm} \intop_{\bar{z}^\pm} 
		\ev{\tilde \phi_A(x^+ \! - \! z^+,\mathbf p_\perp) \tilde \phi_A(x^+ \! - \! \bar z^+,-\mathbf p_\perp)}  \notag\\
		& \qquad \times \ev{\tilde \phi_B(x^- \! - \! z^-,\mathbf q_\perp) \tilde \phi_B(x^- \! - \! \bar z^-,-\mathbf q_\perp)} (\mathbf p \cdot \mathbf q)_\perp^2 J_0(|p+q| \tau_z)J_0(|p+q| \tau_{\bar z}) \notag\\
		&= \frac{g^2}{2} N_c (N_c^2 - 1) \intop_{\mathbf p_\perp,\mathbf q_\perp}  \intop^\infty_0 dZ^+ \intop^\infty_0 dZ^- \intop_{-2Z^+}^{+2Z^+} d\delta z^+ \intop_{-2Z^-}^{+2Z^-} d\delta z^-  (\mathbf p \cdot \mathbf q)_\perp^2 C_A(\mathbf p_\perp) C_B(\mathbf q_\perp)  \notag \\
		& \qquad \times  J_0(|p+q| \tau_z)J_0(|p+q| \tau_{\bar z}) T_A(x^+-Z^+) T_B(x^- -Z^-) U_A(\delta z^+) U_B(\delta z^-).
		\label{eq:e_EL}
	\end{align}
	By combining the results in Eqs.~\eqref{eq:e_BL} and \eqref{eq:e_EL}, the resulting expression for the transverse pressure is given by 
	\begin{align}
		p_T &= \frac{g^2}{2} N_c (N_c^2 - 1) \intop_{\mathbf p_\perp,\mathbf q_\perp}  \intop^\infty_0 dZ^+ \intop^\infty_0 dZ^- \intop_{-2Z^+}^{+2Z^+} d\delta z^+ \intop_{-2Z^-}^{+2Z^-} d\delta z^-  \mathbf p^2_\perp \mathbf q^2_\perp C_A(\mathbf p_\perp) C_B(\mathbf q_\perp)  \notag \\
		& \qquad \times T_A(x^+-Z^+) T_B(x^- -Z^-) U_A(\delta z^+) U_B(\delta z^-)J_0(|p+q| \tau_z)J_0(|p+q| \tau_{\bar z}).
	\end{align}
	This is the main result of this section which shows the dependence of the transverse pressure on the longitudinal structure of the colliding nuclei. We also note that by regularizing the color potential in the auxiliary source terms (see Eq.~\eqref{eq:S_dummy}) as $\phi_{A/B}(x^\pm,\xperp)\equiv \delta(x^\pm)\phi_{A/B}(\mathbf x_\perp)$, the result for 3+1D reduces to the result for 2+1D \cite{Kovner:1995ja,Dumitru:2001ux,McLerran:2016snu}
	\begin{align}
		p_T=\frac{g^2}{2}N_c(N_c^2-1)\intop_{\mathbf p_\perp,\mathbf q_\perp} C_A(\mathbf p_\perp) C_B(\mathbf q_\perp)J_0^2(|p+q| \tau)\mathbf p_\perp^2 \mathbf q_\perp^2.\label{eq:pt_2+1D}
	\end{align}

	\section{Numerical results and comparisons to 3+1D simulations}\label{Sec3}
	
	Basic features of the reaction dynamics for 3+1D collisions have already been examined in detail using real time lattice simulations~\cite{Ipp:2018hai,Schlichting:2020wrv}. In this section we determine the effectiveness of our analytical calculation based on the weak-field approximation by comparing them with full 3+1D simulations. In the 3+1D simulations performed in Ref.~\cite{Schlichting:2020wrv}, we defined the color charge density in Minkowski space as $ \rho^a(t,z,\mathbf x_\perp)=\rho^{a}_{(2D)}(\mathbf x_\perp)T_{R_\gamma}(t+z)$ (for a left moving nucleus),
	which, due to the factorized $\mathbf x_\perp$ and $t + z$ dependence, leads to a less general model for the color charge densities
	\begin{align}
		\ev{\rho^a(t,z, \mathbf x_\perp) \rho^b(t',z', \mathbf x'_\perp) }_\mathrm{3D} = g^2 \overline{\mu}^2 \delta^{ab}  T_{R_\gamma}(t+ z)  T_{R_\gamma}(t'+ z') \GMV(\mathbf x_\perp-\mathbf x'_\perp). \label{eq:corr_3d(tz)}
	\end{align}
	Since the color charges are assumed to be $x^-$ independent, we can write the above two-point function in light cone coordinates as
	\begin{align}
		\ev{\rho^a(x^+, \mathbf x_\perp) \rho^b(x'^+, \mathbf x'_\perp) }_\mathrm{3D} = g^2 \mu'^2\delta^{ab} T_{R'}(x^+) T_{R'}(x'^+) \GMV(\mathbf x_\perp-\mathbf x'_\perp), \label{eq:corr_3d}
	\end{align}
	where 
	\begin{align}
		\mu' = \frac{\overline{\mu}}{\sqrt{2}}, \\
		R' = \frac{R_\gamma}{\sqrt{2}}.
	\end{align}
	Now by comparing Eqs.~\eqref{eq:corr_wf} and \eqref{eq:corr_3d}, one finds that the correlators can be matched by equating the factorized longitudinal dependence as 
	\begin{align}
		\mu^2 T_R(\frac{x^++x'^+}{2}) U_\xi(x^+-x'^+) = \mu'^2 T_{R'}(x^+) T_{R'}(x'^+). 
	\end{align}
	By multiplying the two Gaussians on the left, we find that for $\xi=2R$, the cross terms cancel and then the resulting relations are given as
	\begin{align}
		\mu=\frac{\overline{\mu}}{\sqrt{2}},\\
		R=\frac{R_\gamma}{2}.
	\end{align}
	We can perform an analogous matching for the nuclear model used in Refs.~\cite{Ipp:2017lho, Ipp:2018hai}.  Table~\ref{Table:Parameters} summarizes the parameters for the analytical results obtained from our perturbative expansion (Dilute) and two different 3+1D simulation schemes (3+1D CYM~\cite{Schlichting:2020wrv} and 3+1D CPIC~\cite{Ipp:2017lho, Ipp:2018hai}) with which we determine the extent to which the results of our weak-field approximations agree with the fully non-perturbative real time lattice simulations. Note that $\xi = 2 R$, which we call the coherent limit, is the upper physical limit for the correlation length of color structures inside the nucleus.

	\begin{table}
		\centering
		\begin{tabular}{|l|l|l|}
			\hline
			~$\textbf{Dilute}$      &~$\textbf{3+1D CYM}$ \cite{Schlichting:2020wrv}  &~$\textbf{3+1D CPIC}$ \cite{Ipp:2017lho, Ipp:2018hai}  \\\hline 
			~$g^2\mu$~     &~~~$\frac{g^2\bar{\mu}}{\sqrt{2}}$ ~~& ~~~~$g^2\mu$~ \\
			~$R$~   &~~~$\frac{R_\gamma}{2}$ ~~  & ~~~~$\frac{L}{2}$~\\
			~$m/\Lambda$~ & ~~~$m/\Lambda$ ~~ & ~~~~$m/\Lambda$~\\
			\hline
		\end{tabular}
		\caption{Parameters for comparing (semi-)analytical results with 3+1D simulations.}
		\label{Table:Parameters}
	\end{table}

\begin{figure}
		\centering
		\includegraphics[width=0.99\textwidth]{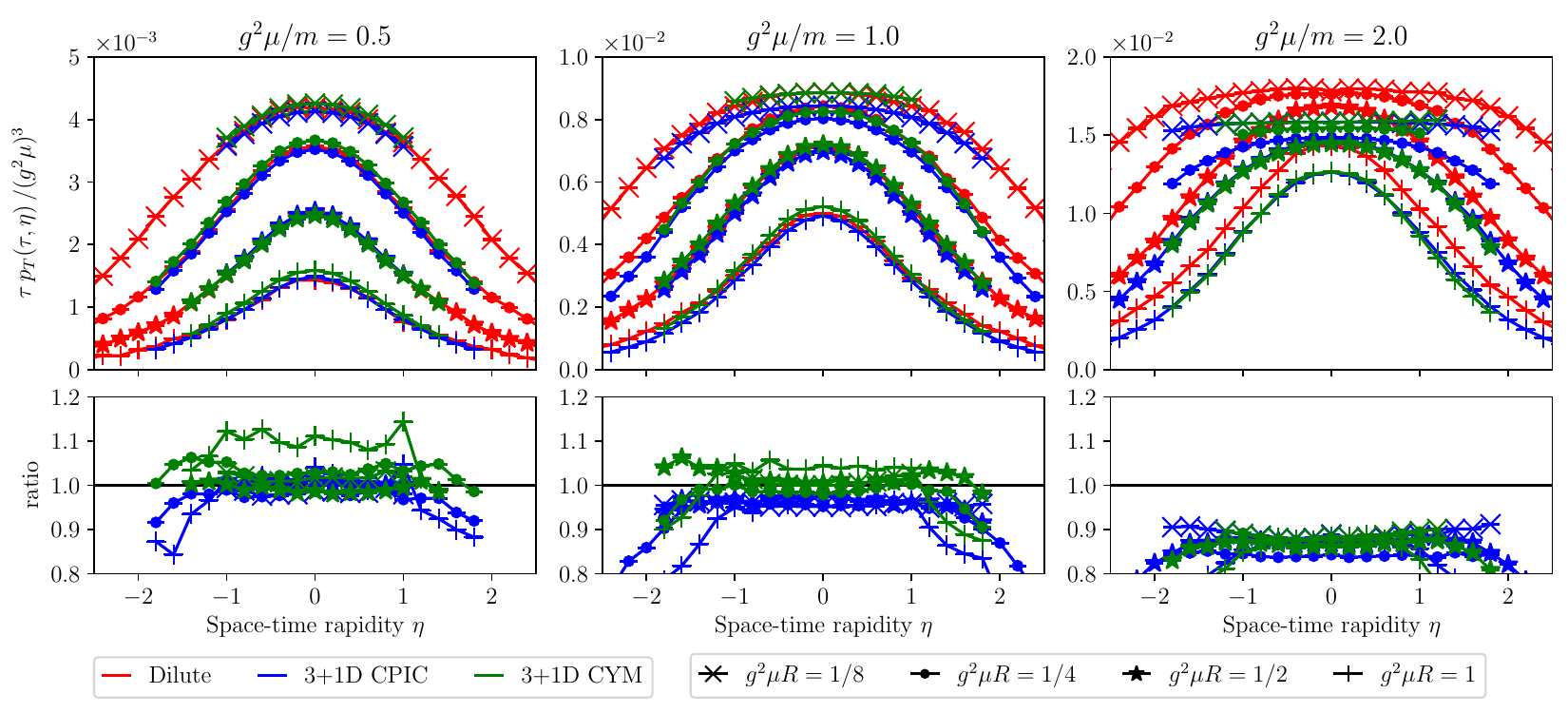}
		\caption{Transverse pressure as a function of rapidity for three different simulation parameters in the dilute limit: $g^2\mu/m=0.5$ (left), $g^2\mu / m=1.0$ (middle) and $g^2\mu/m=2.0$ (right) compared to results from non-perturbative classical Yang-Mills simulations (3+1D CPIC \cite{Ipp:2017lho, Ipp:2018hai} and 3+1D CYM \cite{Schlichting:2020wrv}). 
		}
		\label{fig:MC_vs_full3Dsimulations}
	\end{figure}

	We note that the non-linearity, which measures the strength of diluteness of a model, can be controlled by the dimensionless ratio of the color charge density $g^2\mu$ and the infrared regulator $m$. We vary this dimensionless parameter and compare the transverse pressure as obtained from the analytical result and the result from 3+1D simulations, for different longitudinal extents of the colliding nuclei in Fig.~\ref{fig:MC_vs_full3Dsimulations}. 
	We choose the same transverse lattice discretization for both schemes:
	$ma_\perp=0.125$ and $m a_{\perp} N_{\perp}=16$ with $\Lambda/m=5$. Since the two simulations rely on completely different numerical schemes,\footnote{Essential differences regard the lattice discretization of the color currents. Within the colored particle-in-cell formalism (3+1D CPIC) of~\cite{Ipp:2018hai}, eikonal color currents are propagated as colored particles and not subject to a lattice dispersion, whereas in the 3+1D CYM formalism of \cite{Schlichting:2020wrv} eikonal currents are propagated based on the current conservation equations and subject to the lattice dispersion. \vspace{4em}} the longitudinal discretization of the lattice is different, with $R_\gamma / a_z = 16$ in all 3+1D CYM simulations and $R_\gamma / a_z \in \{8, 16, 32 \}$ in the case of 3+1D CPIC simulations. In both cases, the discretization is chosen such that the nuclei are properly resolved $R_\gamma/a_z\gg 1$, and the longitudinal extent $N_z a_z$ is large enough to allow for sufficiently long simulation times, i.e.~$N_z a_z\gg R_{\gamma}$. The results for the dilute approximation are computed using Monte Carlo integration.
	
	We primarily focus our attention on coherent color fields $\xi=2 R$ for which the nuclear model used for analytical calculations and simulations is identical. Since we are interested in late times, where in the boost-invariant limit the transverse pressure per unit rapidity $\tau p_{T}(\tau)$ becomes independent of the proper-time $\tau$, we scale the transverse pressure in Fig.~\ref{fig:MC_vs_full3Dsimulations} with proper time $\tau$ to eliminate the leading time dependence. Numerical results in the dilute approximation are evaluated at $m \tau = 8$, while the 3+1D simulations are evaluated at $g^2 \mu \tau = 2$ for $g^2 \mu R = 1/2, 1$ and all values of $g^2 \mu / m$, whereas for $g^2 \mu R = 1/8, 1/4$ we set $g^2 \mu \tau = 1$ for all $g^2 \mu / m$, except for $g^2 \mu / m = 2$, $g^2 \mu R = 1/4$ where we use $g^2 \mu \tau = 2$.\footnote{Note that for a proper comparison, we interpolate data on the $(t,z)$ grids of the 3+1D simulations before switching to $(\tau, \eta)$ coordinates. To reduce statistical fluctuations, we use rapidity bins of width $\Delta \eta = 0.2$.} Before discussing the results of our weak field approximation, we emphasize that the results of the two different 3+1D classical Yang-Mills
	implementations~(3+1D CYM~\cite{Schlichting:2020wrv} and 3+1D CPIC~\cite{Ipp:2017lho, Ipp:2018hai})  are in excellent agreement with each other. We find that, as per our expectation, the analytical calculation works remarkably well in the dilute limit $g^2\mu/m=0.5$ as seen from the left panel of Fig.~\ref{fig:MC_vs_full3Dsimulations}. By increasing the non-linearity of the model $g^2\mu/m\geq1$, we find that the analytical results in the dilute limit overestimate the transverse pressure; nevertheless the rapidity profiles are still reproduced rather well and the flattening of the rapidity profiles with increasing $g^2\mu/m$ is correctly predicted by the (semi-)analytic calculation.\footnote{It should be noted that we use the dimensionless length scale $g^2 \mu R$ for the sake of comparing the weak field approximation to our non-perturbative simulations. However in the dilute limit, $g^2 \mu R$ is in fact not a particularly useful scale, because $g^2 \mu$ only enters as an overall normalization factor of $T^{\mu\nu}$. The more appropriate length scale is given by $m R$, i.e.~when $mR$ and $m/\Lambda$ are fixed, the shape of the rapidity profile does not change with the non-linearity parameter $g^2 \mu /m$. The widening of the dilute rapidity profiles in Fig.~\mbox{\ref{fig:MC_vs_full3Dsimulations}} for fixed $g^2 \mu R$ and increasing $g^2 \mu / m$ should be interpreted as widening due to varying $m R$.}
	It is further interesting to note that the ratio of the analytical to that of the simulation results are roughly the same for different thicknesses of the colliding nuclei, which suggests that the non-linearity could effectively be introduced by re-scaling the pressure profile. Besides the significantly smaller computational cost, another enormous benefit of the semi-analytic calculation is that it is not bounded by lattice size, and therefore, we are able to perform computations at larger rapidities, as is clearly visible from Fig.~\ref{fig:MC_vs_full3Dsimulations}.
	
	\begin{figure*}
		\centering
		\includegraphics[width=0.48\textwidth]{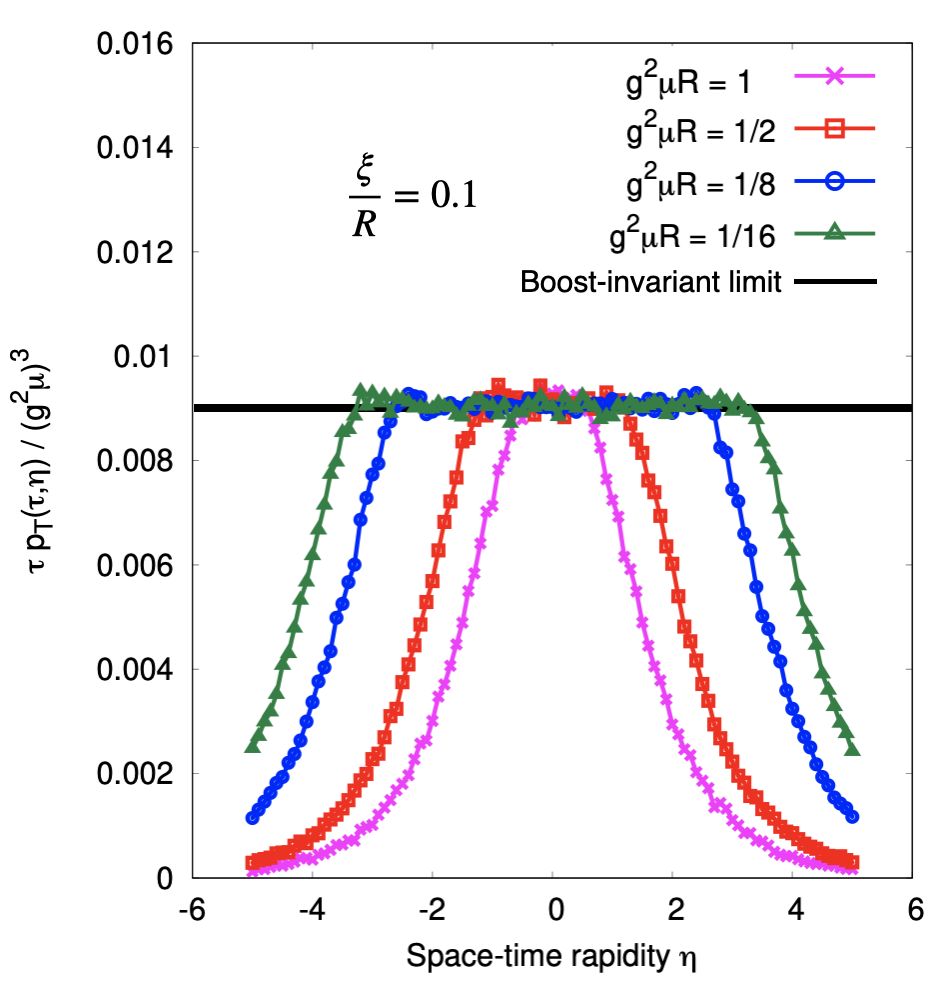}
		\hfill
		\includegraphics[width=0.48\textwidth]{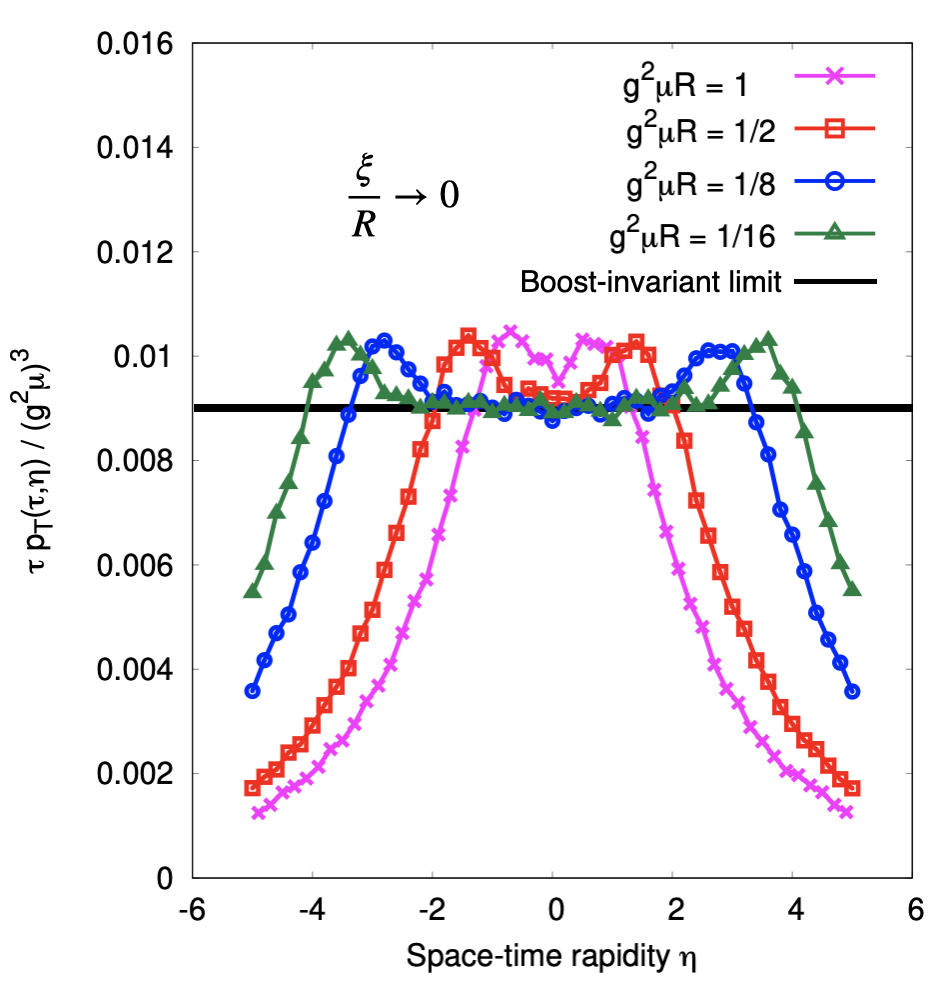}
		\\
		\includegraphics[width=0.48\textwidth]{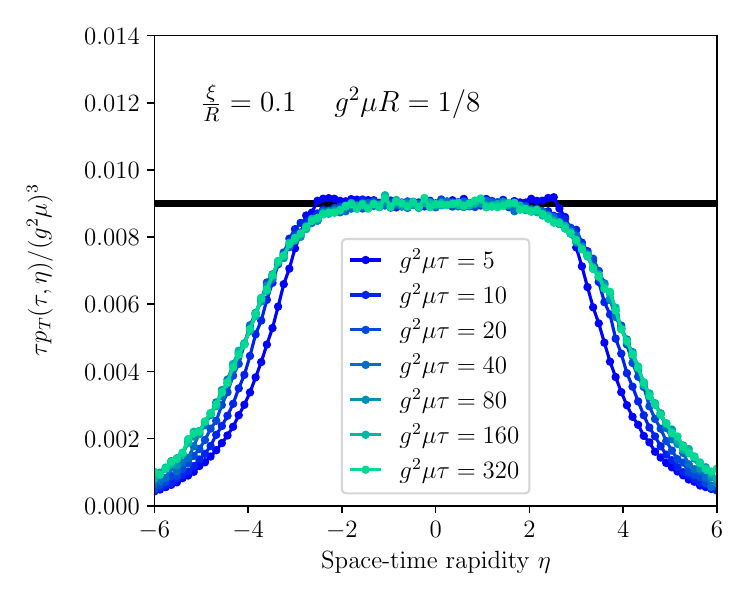}
		\hfill
		\includegraphics[width=0.48\textwidth]{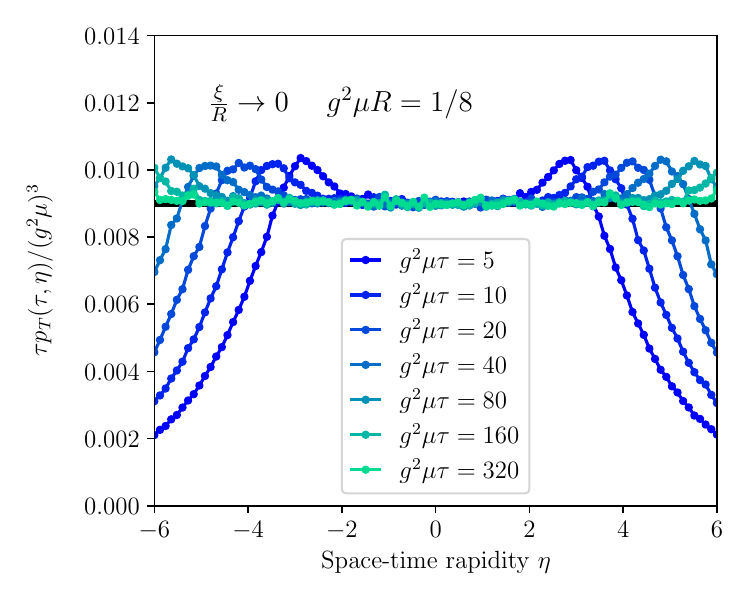}
		\caption{Transverse pressure corresponding to two different limits of coherence length: $\xi/R = 0.1$ (left) and $\xi/R \rightarrow0$ (right), for varying thickness of the incoming nuclei (top plots) and varying proper time (bottom plots).
		}
		\label{fig:epsbysigma_limit}
	\end{figure*}
	
	Having established that our analytical expressions reproduce the full 3+1D numerical simulations in the dilute limit, we consider the various limits of our nuclear model. We start by looking at the coherence length $\xi/R$, which accounts for the randomness of color charges across a fixed longitudinal extent of the nucleus. Naturally, the longitudinal extent of the nucleus ($R$) is greater than the 
	size of a correlated region within the nucleus ($\xi$)
	and hence for a physical limit $\xi/R \lesssim 1$. In Fig.~\ref{fig:epsbysigma_limit}, we plot the transverse pressure for different thicknesses of colliding nuclei while considering two different values of $\xi/R = 0.1$ (left panels), which roughly corresponds to the ratio $A^{-1/3}$ of the size of a nucleon and a nucleus in large nuclei, and $\xi/R \rightarrow 0$ (right panels), which corresponds to the McLerran-Venugopalan (MV) model $(A\to \infty)$. For these plots we use $m = g^2 \mu$ and $\Lambda / m = 5$. The top plots show the profiles for different $g^2\mu R$ at fixed time $g^2 \mu \tau = 5$, whereas the bottom plots are evaluated for fixed $g^2\mu R=1/8$ varying $g^2 \mu \tau$. We further include a comparison of the results of the 3+1D dilute calculation to the corresponding result in the 2+1D boost-invariant limit, which is obtained by integrating Eq.~\eqref{eq:pt_2+1D}. We observe that the profiles approach the same boost-invariant plateau around mid-rapidity, whereas the flanks at larger rapidities are different and depend on the correlation length $\xi/R$. By decreasing the thickness of the colliding nuclei $g^2\mu R\rightarrow 0$, one approaches the boost-invariant limit, where the central plateau extends across larger and larger rapidity intervals. With regards to the proper time dependence, we find that in the limit $\xi / R \rightarrow 0$ in Fig.~\ref{fig:epsbysigma_limit} the profiles exhibit a significant time dependence up to very late times, where the flanks continue to move towards larger rapidities, while the central plateau remains time-independent. In contrast, we observe that the time dependence is much milder for $\xi / R = 0.1$ (Fig.~\ref{fig:epsbysigma_limit} on the bottom left), where a stable profile is reached quickly and the double-peak structure vanishes entirely. 
	
	\begin{figure*}
		\centering
		\includegraphics[width=0.44\textwidth]{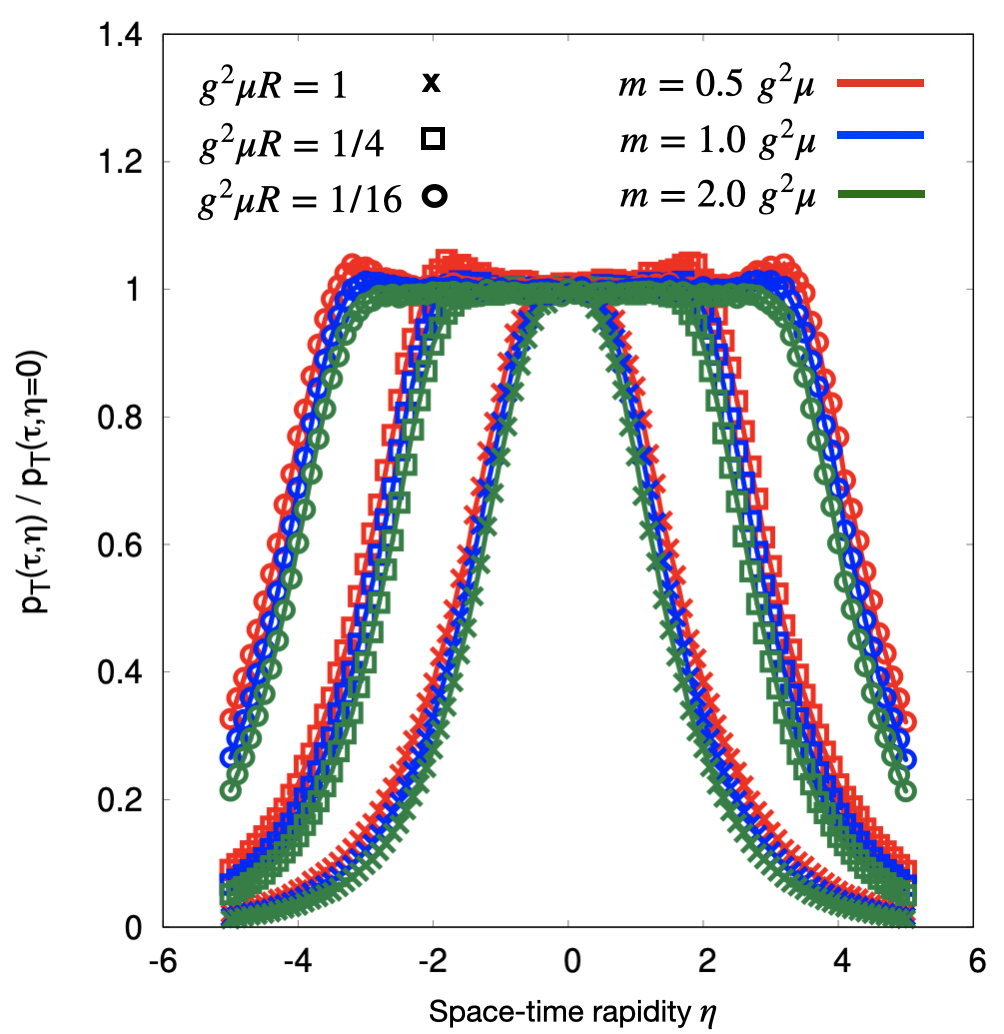}
		\includegraphics[width=0.44\textwidth]{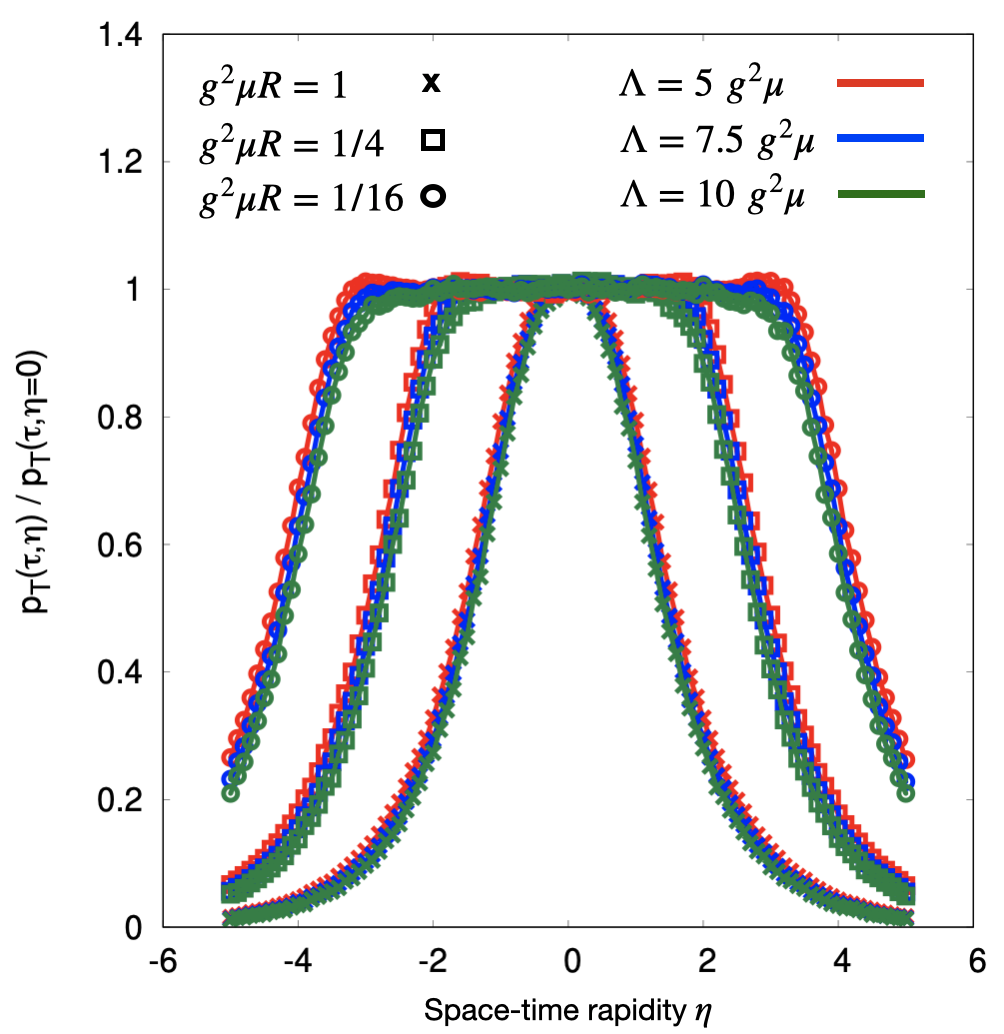}
		\caption{Transverse pressure relative to its value at mid-rapidity for varying thickness of colliding nuclei along with fixed UV regulator (left) and fixed infrared regulator (right).
			\label{fig:UV_and_IR}
		}
	\end{figure*}
	
	Finally, we investigate the dependence of the rapidity profile on the UV  and IR regulators. In Fig.~\ref{fig:UV_and_IR} we plot the transverse pressure normalized to its value at mid-rapidity for different longitudinal extents $g^2\mu R$ and $\xi/R=0.1$. In the left panel we fix the UV regulator to $\Lambda=5g^2\mu$ and vary the infrared regulator $m$ to three different values. Similarly, for the right panel the IR regulator is set to constant $m=g^2\mu$, and $\Lambda$ takes three different values. We observe that for a constant proper time $g^2\mu \tau=5$, the profiles are largely insensitive to the variation  apart from small deviations in the shoulders and flanks. 
	Similar to  Fig.~\ref{fig:epsbysigma_limit}, a boost-invariant plateau around mid-rapidity emerges upon decreasing the thickness of the colliding nuclei and the width of the plateau appears to be insensitive to the UV and IR regulators. 
	
	\section{Conclusions and outlook} \label{Sec4}
	
	We performed the first analytic calculation of the longitudinal profiles of the energy deposition in heavy-ion collisions 
	within the dilute limit of the Color Glass Condensate effective field theory of high-energy QCD. We obtained general analytic expressions for the color fields of the Glasma produced in the future light cone (cf. Eqs.~\eqref{eq:ap_final} -- \eqref{eq:ai_final}), and employed them to study the rapidity dependence of the transverse pressure for a simplified nuclear model including non-trivial longitudinal color correlations.
	
	By comparing the (semi-)analytic results in the dilute approximation to non-perturbative 3+1D classical Yang-Mills simulations, we confirm excellent agreement in the dilute regime. Even beyond the dilute limit, our approximation appears to capture the rapidity profiles rather well, while the overall magnitude of energy deposition is overestimated, once non-linear effects become important. 
	
	Since our analytic expressions allow for an efficient numerical determination of the energy momentum tensor $T^{\mu\nu}$, the results presented in this paper provide new opportunities to further explore the longitudinal structure of matter produced in high-energy heavy-ion collisions, to study e.g.~the interplay of longitudinal and transverse fluctuations and develop new Monte Carlo event generators for the initial state of heavy-ion collisions.

	\textit{Acknowledgement:} 
	We thank B.~Schenke and T.~Lappi for discussions and collaboration on related projects.
	DM is supported by the Austrian Science Fund FWF No.~P32446-N27.
	SS and PS are supported under the Deutsche Forschungsgemeinschaft (DFG, German Research Foundation) through the CRC-TR 211 `Strong-interaction matter under extreme conditions' - project number: 315477589 TRR-211. The computations in this work were performed at the Paderborn Center for Parallel Computing (PC2) and the Vienna Scientific Cluster (VSC).
	
	\bibliographystyle{elsarticle-num}
	\bibliography{bib}

\end{document}